\input harvmac.tex

\vskip 1.5in
\Title{\vbox{\baselineskip12pt
\hbox to \hsize{\hfill}
\hbox to \hsize{\hfill WITS-CTP-073}}}
{\vbox{
	\centerline{\hbox{A String Model for AdS Gravity
		}}\vskip 5pt
        \centerline{\hbox{and Higher Spins
		}} } }
\centerline{Dimitri Polyakov\footnote{$^\dagger$}
{dimitri.polyakov@wits.ac.za}}
\medskip
\centerline{\it National Institute for Theoretical Physics (NITHeP)}
\centerline{\it  and School of Physics}
\centerline{\it University of the Witwatersrand}
\centerline{\it WITS 2050 Johannesburg, South Africa}

\vskip .3in

\centerline {\bf Abstract}

We construct a string sigma-model which low energy limit
describes the anti de Sitter gravity and spin 3 massless
fields in Vasiliev's frame-like
formalism. The model is
based on vertex operators generating vielbein and connection fields
in the Mac Dowell - Mansoury - Stelle - West (MMSW) formulation of gravity.
The structure  of the vertex operators  is based on the hidden symmetry 
generators in RNS superstring theory, realizing the isometry
group of the AdS space.
The beta-function equations in the sigma-model lead to 
equations of motion in the MMSW gravity with negative cosmological constant
with the AdS geometry being the vacuum solution.
Generalizations for the higher spin fields are analyzed and 
equations of motion for spin 3 fields in $d=3$ in frame-like 
formalism   are obtained in the low energy limit of string theory.
These equations correspond to those of $sl(3,R)$ truncated
Chern-Simons action based on higher spin algebra $hs(1,1)$.

\Date{June 2011}

\vfill\eject

\lref\fvf{E.S. Fradkin, M.A. Vasiliev, Nucl. Phys. B 291, 141 (1987)}
\lref\fvs{E.S. Fradkin, M.A. Vasiliev, Phys. Lett. B 189 (1987) 89}
\lref\mmswf{S.W. MacDowell, F. Mansouri, Phys. Rev.Lett. 38 (1977) 739}
\lref\mmsws{K. S. Stelle and P. C. West, Phys. Rev. D 21 (1980) 1466}
\lref\mmswt{C.Preitschopf and M.A.Vasiliev, hep-th/9805127}
\lref\vmaf{M. A. Vasiliev, Sov. J. Nucl. Phys. 32 (1980) 439,
Yad. Fiz. 32 (1980) 855}
\lref\vmas{V. E. Lopatin and M. A. Vasiliev, Mod. Phys. Lett. A 3 (1988) 257}
\lref\vmat{E.S. Fradkin and M.A. Vasiliev, Mod. Phys. Lett. A 3 (1988) 2983}
\lref\vmafth{M. A. Vasiliev, Nucl. Phys. B 616 (2001) 106 }
\lref\bianchi{ M. Bianchi, V. Didenko, arXiv:hep-th/0502220}
\lref\ruehl{R. Manvelyan and W. Ruehl, hep-th/0502123}
\lref\bonellio{G. Bonelli, JHEP 0411 (2004) 059}
\lref\hsaone{M. A. Vasiliev, Fortsch. Phys. 36 (1988) 33}
\lref\hsatwo{E.S. Fradkin and M.A. Vasiliev, Mod. Phys. Lett. A 3 (1988) 2983}
\lref\hsathree{S. E. Konstein and M. A. Vasiliev, Nucl. Phys. B 331 (1990) 475}
\lref\hsafour{M.P. Blencowe, Class. Quantum Grav. 6, 443 (1989)}
\lref\hsafive{E. Bergshoeff, M. Blencowe and K. Stelle, 
Comm. Math. Phys. 128 (1990) 213}
\lref\hsasix{E. Sezgin and P. Sundell, Nucl. Phys. B 634 (2002) 120 }
\lref\hsaseven{M. A. Vasiliev, Phys. Rev. D 66 (2002) 066006 }
\lref\soojongf{M. Henneaux, S.-J. Rey, JHEP 1012:007,2010}
\lref\henneaux{J. D. Brown and M. Henneaux, Commun. Math. Phys. 104, 207 (1986)}
\lref\sagnottia{A. Sagnotti, E. Sezgin, P. Sundell, hep-th/0501156}
\lref\sorokin{D. Sorokin, AIP Conf. Proc. 767, 172 (2005)}
\lref\fronsdal{C. Fronsdal, Phys. Rev. D18 (1978) 3624}
\lref\coleman{ S. Coleman, J. Mandula, Phys. Rev. 159 (1967) 1251}
\lref\haag{R. Haag, J. Lopuszanski, M. Sohnius, Nucl. Phys B88 (1975)
257}
\lref\weinberg{ S. Weinberg, Phys. Rev. 133(1964) B1049}
\lref\tseytbuch{E. Buchbinder, A. Tseytlin, JHEP 1008:057,2010}
\lref\fradkin{E. Fradkin, M. Vasiliev, Phys. Lett. B189 (1987) 89}
\lref\skvortsov{E. Skvortsov, M. Vasiliev, Nucl.Phys.B756:117-147 (2006)}
\lref\skvortsovb{E. Skvortsov, J.Phys.A42:385401 (2009)}
\lref\mva{M. Vasiliev, Phys. Lett. B243 (1990) 378}
\lref\mvb{M. Vasiliev, Int. J. Mod. Phys. D5
(1996) 763}
\lref\mvc{M. Vasiliev, Phys. Lett. B567 (2003) 139}
\lref\brink{A. Bengtsson, I. Bengtsson, L. Brink, Nucl. Phys. B227
 (1983) 31}
\lref\deser{S. Deser, Z. Yang, Class. Quant. Grav 7 (1990) 1491}
\lref\bengt{ A. Bengtsson, I. Bengtsson, N. Linden,
Class. Quant. Grav. 4 (1987) 1333}
\lref\boulanger{ X. Bekaert, N. Boulanger, S. Cnockaert,
J. Math. Phys 46 (2005) 012303}
\lref\metsaev{ R. Metsaev, arXiv:0712.3526}
\lref\siegel{ W. Siegel, B. Zwiebach, Nucl. Phys. B282 (1987) 125}
\lref\siegelb{W. Siegel, Nucl. Phys. B 263 (1986) 93}
\lref\nicolai{ A. Neveu, H. Nicolai, P. West, Nucl. Phys. B264 (1986) 573}
\lref\damour{T. Damour, S. Deser, Ann. Poincare Phys. Theor. 47 (1987) 277}
\lref\sagnottib{D. Francia, A. Sagnotti, Phys. Lett. B53 (2002) 303}
\lref\sagnottic{D. Francia, A. Sagnotti, Class. Quant. Grav.
 20 (2003) S473}
\lref\sagnottid{D. Francia, J. Mourad, A. Sagnotti, Nucl. Phys. B773
(2007) 203}
\lref\labastidaa{ J. Labastida, Nucl. Phys. B322 (1989)}
\lref\labastidab{ J. Labastida, Phys. Rev. Lett. 58 (1987) 632}
\lref\mvd{L. Brink, R.Metsaev, M. Vasiliev, Nucl. Phys. B 586 (2000) 183}
\lref\klebanov{ I. Klebanov, A. M. Polyakov,
Phys.Lett.B550 (2002) 213-219}
\lref\mve{
X. Bekaert, S. Cnockaert, C. Iazeolla,
M.A. Vasiliev,  IHES-P-04-47, ULB-TH-04-26, ROM2F-04-29, 
FIAN-TD-17-04, Sep 2005 86pp.}
\lref\sagnottie{A. Campoleoni, D. Francia, J. Mourad, A.
 Sagnotti, Nucl. Phys. B815 (2009) 289-367}
\lref\sagnottif{
A. Campoleoni, D. Francia, J. Mourad, A.
 Sagnotti, arXiv:0904.4447}
\lref\sagnottig{D. Francia, A. Sagnotti, J.Phys.Conf.Ser.33:57 (2006)}
\lref\selfa{D. Polyakov, Int.J.Mod.Phys.A20:4001-4020,2005}
\lref\selfb{ D. Polyakov, arXiv:0905.4858}
\lref\selfc{D. Polyakov, arXiv:0906.3663, Int.J.Mod.Phys.A24:6177-6195 (2009)}
\lref\selfd{D. Polyakov, Phys.Rev.D65:084041 (2002)}
\lref\spinself{D. Polyakov, Phys.Rev.D82:066005,2010}
\lref\spinselff{D. Polyakov,Phys.Rev.D83:046005,2011}
\lref\mirian{A. Fotopoulos, M. Tsulaia, Phys.Rev.D76:025014,2007}
\lref\extraa{I. Buchbinder, V. Krykhtin,  arXiv:0707.2181}
\lref\extrab{I. Buchbinder, V. Krykhtin, Phys.Lett.B656:253-264,2007}
\lref\extrac{X. Bekaert, I. Buchbinder, A. Pashnev, M. Tsulaia,
Class.Quant.Grav. 21 (2004) S1457-1464}
\lref \extrad{I. Buchbinder, A. Pashnev, M. Tsulaia,
arXiv:hep-th/0109067}
\lref\extraf{I. Buchbinder, A. Pashnev, M. Tsulaia, 
Phys.Lett.B523:338-346,2001}
\lref\extrag{I. Buchbinder, E. Fradkin, S. Lyakhovich, V. Pershin,
Phys.Lett. B304 (1993) 239-248}
\lref\extrah{I. Buchbinder, A. Fotopoulos, A. Petkou, 
 Phys.Rev.D74:105018,2006}
\lref\bonellia{G. Bonelli, Nucl.Phys.B {669} (2003) 159}
\lref\bonellib{G. Bonelli, JHEP 0311 (2003) 028}
\lref\ouva{C. Aulakh, I. Koh, S. Ouvry, Phys. Lett. 173B (1986) 284}
\lref\ouvab{S. Ouvry, J. Stern, Phys. Lett.  177B (1986) 335}
\lref\ouvac{I. Koh, S. Ouvry, Phys. Lett.  179B (1986) 115 }
\lref\hsself{D.Polyakov, arXiv:1005.5512}
\lref\sundborg{ B. Sundborg, ucl.Phys.Proc.Suppl. 102 (2001)}
\lref\sezgin{E. Sezgin and P. Sundell,
Nucl.Phys.B644:303- 370,2002}
\lref\morales{M. Bianchi,
J.F. Morales and H. Samtleben, JHEP 0307 (2003) 062}
\lref\giombif{S. Giombi, Xi Yin, arXiv:0912.5105}
\lref\giombis{S. Giombi, Xi Yin, arXiv:1004.3736}
\lref\bekaert{X. Bekaert, N. Boulanger, P. Sundell, arXiv:1007.0435}
\lref\taronna{A. Sagnotti, M. Taronna, arXiv:1006.5242, 
Nucl.Phys.B842:299-361,2011}
\lref\zinoviev{Yu. Zinoviev, arXiv:1007.0158}
\lref\fotopoulos{A. Fotopoulos, M. Tsulaia, arXiv:1007.0747}
\lref\fotopouloss{A. Fotopoulos, M. Tsulaia, arXiv:1009.0727}
\lref\taronnao{M. Taronna, arXiv:1005.3061}
\lref\taronnas{A. Sagnotti, M. Taronna, arXiv:1006.5242 ,
Nucl.Phys.B842:299-361,2011}
\lref\campo{A.Campoleoni,S. Fredenhagen,S. Pfenninger, S. Theisen,
arXiv:1008.4744, JHEP 1011 (2010) 007}
\lref\gaber{M. Gaberdiel, T. Hartman, arXiv:1101.2910, JHEP 1105 (2011) 031}
\lref\per{	
N. Boulanger,S. Leclercq, P. Sundell, JHEP 0808(2008) 056 }
\lref\mav{V. E. Lopatin and M. A. Vasiliev, Mod. Phys. Lett. A 3 (1988) 257}
\lref\zinov{Yu. Zinoviev, Nucl. Phys. B 808 (2009)}
\lref\sv{E.D. Skvortsov, M.A. Vasiliev,
Nucl. Phys.B 756 (2006)117}
\lref\mvasiliev{D.S. Ponomarev, M.A. Vasiliev, Nucl.Phys.B839:466-498,2010}
\lref\zhenya{E.D. Skvortsov, Yu.M. Zinoviev, arXiv:1007.4944}
\lref\perf{N. Boulanger, C. Iazeolla, P. Sundell, JHEP 0907 (2009) 013 }
\lref\pers{N. Boulanger, C. Iazeolla, P. Sundell, JHEP 0907 (2009) 014 }
\lref\selft{D. Polyakov,Phys.Rev.D82:066005,2010}
\lref\selftt{D. Polyakov, Int.J.Mod.Phys.A25:4623-4640,2010}
\lref\tseytlin{I. Klebanov, A Tseytlin, Nucl.Phys.B546:155-181,1999}
\lref\ruben{R. Manvelyan, K. Mkrtchyan, W. Ruehl, arXiv:1009.1054}
\lref\rubenf{R. Manvelyan, K. Mkrtchyan, W. Ruehl, Nucl.Phys.B836:204-221,2010}
\lref\robert{
R. De Mello Koch, A. Jevicki, K. Jin, J. A. P. Rodrigues, arXiv:1008.0633}
\lref\bek{X. Bekaert, S. Cnockaert, C. Iazeolla, M. A. Vasiliev,
hep-th/0503128}
\lref\vcubic{M. Vasiliev, arXiv:1108.5921}
\centerline{\bf  1. Introduction}
Describing higher spin fields in anti-de Sitter geometry
is a fascinating and challenging problem ~{\fvf, \fvs}.
As there is no well-defined S-matrix
in  AdS geometry, one could hope to circumvent
no-go theorems ~{\coleman, \haag} and look for
consistently interacting theories of higher spins.
At the same time,  higher spin fields in AdS backgrounds
are important ingredients of the AdS/CFT correspondences, 
as there are multitudes of corresponding operators appearing
in dual conformal field theories (e.g. see ~{\klebanov}).
Understanding relations between holography and higher spin dynamics
is therefore crucial for the entire concept
of $AdS/CFT$ in general.
A frame-like  formalism is a particularly
powerful and efficient approach
to gravity and higher spin field theories in curved backgrounds
~{\vmaf, \vmas, \vmat, \vmafth, \vcubic}. In this approach,
gravity and higher spin field theories are formulated in terms of
gauge theories of vielbeins and connections. 
 For an ordinary theory of gravity, such an approach has
been first developed by Cartan and Weyl and then generalized by
Mac Dowell, Mansouri, Stelle and West   who proposed
manifestly gauge-invariant frame-like formulation of gravity
with nonvanishing cosmological constant ~{\mmswf, \mmsws}.
 The frame-like approach
for the higher spin fields
is the generalization of the MMSW formalism for fields with spins greater
than 2. It
 has been proposed by Vasiliev ~{\vmaf, \vmas} 
and later developed in a number of papers . In particular, equations
of motion for anti- de Sitter (AdS) gravity , as well as for the higher
spin fields in AdS space-time become remarkably compact and elegant
in this formalism.

String theories in AdS backgrounds constitute, in turn, another crucial
ingredient of the AdS/CFT correspondence. In fact, this correspondence
can be most naturally understood as isomorphism between vertex operators
on the string theory side in AdS space and appropriate 
observables in conformal field theory, so their correlation functions
 match exactly.It is actualy the perturbative dynamics of strings in AdS space
that could provide a powerful test for the AdS/CFT, in order to approach the
strongly coupled regime of gauge theories.
Unfortunately, string theory in AdS backgrounds,
in its standard formulations, is difficult to approach beyond 
semiclassical limit, although even in this limit some remarkable
results for anomalous dimensions of gauge theory operators have been obtained
(e.g. see ~{\tseytbuch}).

At the same time, string theory  appears to be an efficient and natural
framework to describe the dynamics of interacting higher spin field
theories both in flat space and in AdS
~{\sagnottia, \sagnottib, \sagnottic, \sagnottib, 
 \sagnottie, \sagnottif, \sagnottig, \taronnao, \taronnas 
\labastidaa, \labastidab,\metsaev,\ruben, \rubenf, \fotopoulos, \fotopouloss,
\spinself, \spinselff}

In this paper we propose a sigma-model in RNS string theory, based
on hidden symmetry generators ~{\selfc} that realise the $o(d-1,2)$ isometry
algebra of $AdS_d$. Namely, the sigma model is based on the RNS superstring
theory perturbed by the vertex operators which structure is determined
by the AdS isometry  generators. As will be demonstrated below,
the vertex operators, constructed in this work,  can be regarded as sources
for connection gauge fields and vielbeins in space-time (which can be unified
into a single $o(d-1,2)$ connection gauge field).
The construction is based on certain hidden space-time symmetry
also generators (reviewed in this paper )
that realize AdS isometry group in $d$ dimensions.
It is particularly remarkable that the commutation relations of these operators
(computed in the Section 2) fix the negative sign of the 
cosmological constant, leading to the appearance of the AdS  geometry
in the sigma-model constructed in the paper.
 BRST nontriviality 
constraints
on the vertex operators in the sigma-model
 lead to $o(d-1,2)$ gauge transformations on
the unified connection, while the BRST invariance conditions produce
linearized equations of motion for the connection field (including
the zero torsion constraint). Beta-function equations for the sigma-model,
 in turn, lead to full equations of motion for 
AdS gravity (with cosmological constant) in the frame-like formalism.
The rest of the paper is organized as follows.
In the next section (section 2) 
we review the realization of AdS isometry by special matter-ghost
mixing symmetry generators ($\alpha$-symmetries ~{\selfc}). These generators
will be used as building blocks to construct closed string vertex operators
for the unified connection gauge field. 
In Section 3, we analyze the BRST constraints for the connection
vertex operators
 We find that the BRST nontriviality conditions lead to
gauge symmetry transformations for the $o(d-1,2)$ connection  field in
MMSW gravity , while the BRST-invariance conditions lead to linearized
equations of motion for this field.
In Section 4, we study the sigma-model in RNS string theory, based on the
constructed vertex operators for the connection.
We calculate the beta-function in this model and show
that it leads to full equations of motion for 
MMSW gravity in the frame-like formalism with cosmological constant.
The source of the cosmological constant comes from the vertices corresponding
to transvections in AdS isometry transformations.

In the concluding section we discuss the physical implications of our results
and their generalizations to higher spin field theories in the frame-like 
approach, particularly deriving frame-like equations of motion
for spin 3 fields on $AdS_3$ from string theory sigma-model.

\centerline{\bf 2. AdS Isometry and Space-Time $\alpha$-Symmetry }

In string theory the space-time symmetry generators are typically
conformal dimension 1 primary fields, integrated over the worldsheet 
boundary. For example, in RNS string theory the Poincare isometries
of flat space-time are realized by the operators of translations and rotations
given by
\eqn\grav{\eqalign{
T_m=\oint{{dz}\over{2i\pi}}\partial{X^m}\cr
L_{mn}=\oint{{dz}\over{2i\pi}}\psi_{m}\psi_n+...}}
where we have skipped the ghost dependent terms in the expression
for the rotation generator (that ensure the overall BRST invariance
of the generator). Here $X^m (m=0,...,d-1)$  are the space-time coordinates,
and $\psi^m$ are their worldsheet superpartners.
What is far less trivial is that, in addition
the standard Poincare isometries, RNS string theory also possesses
a set of additional surprising symmetries that are realized nonlinearly
and mix matter and ghost degrees of freedom ~{\selfc}.
In particular , there is a subgroup of these generators that realize
the $o(d-1,2)$ isometry of $AdS_d$.
Recall that the $AdS_d$ isometry algebra is given by:
\eqn\grav{\eqalign{
{\lbrack}T_{ab},T_{cd}\rbrack
=\eta_{ac}T_{bd}-\eta_{ab}T_{cd}-\eta_{cd}T_{ab}+\eta_{bd}T_{ac}\cr
{\lbrack}T_{a},T_{bc}\rbrack
=\eta_{ab}T_{c}-\eta_{ac}T_{b}\cr
{\lbrack}T_{a},T_{b}\rbrack
=\Lambda{T_{ab}}}}
where $\Lambda\sim{-}{1\over{\rho^2}}$
is negative cosmological constant and $R$ is the $AdS$ radius.
In other words, the main property distinguishing the $AdS$ 
isometry algebra (2)
from the one of the flat space is the noncommutation of the vector generators
(proportional to the cosmological constant). For this reason, these generators
are known as the generators of transvections in the AdS space, 
to distinguish them from translations in flat space-time .

In RNS string theory, the AdS isometry algebra (2) can be realized by using the 
the generators of the $\alpha$-symmetries ~{\selfc} inducing 
nonlinear global symmetries in space-time.
Namely, consider the RNS superstring theory in flat space with the
 action given by:
\eqn\grav{\eqalign{S_{RNS}=S_{matter}+S_{bc}+S_{\beta\gamma}+S_{Liouville}\cr
S_{matter}=-{1\over{4\pi}}\int{d^2z}(\partial{X_m}\bar\partial{X^m}
+\psi_m\bar\partial\psi^m+{\bar\psi}_m\partial{\bar\psi}^m)\cr
S_{bc}={1\over{2\pi}}\int{d^2z}(b\bar\partial{c}+{\bar{b}}\partial
{\bar{c}})\cr
S_{\beta\gamma}={1\over{2\pi}}\int{d^2z}(\beta\bar\partial\gamma
+\bar\beta\partial{\bar\gamma})
\cr
S_{Liouville}=-{1\over{4\pi}}\int{d^2z}(\partial\varphi\bar\partial\varphi
+\bar\partial\lambda\lambda+\partial\bar\lambda\bar\lambda
+\mu_0{e^{B\varphi}}(\lambda\bar\lambda+F))
}}
where $\varphi,\lambda, F$ are components of super Liouville field
and the Liouville background charge is
\eqn\lowen{
q=B+B^{-1}={\sqrt{{{9-d}\over2}}}}

The ghost fields $b,c,\beta,\gamma$ bosonized according to

\eqn\grav{\eqalign{b=e^{-\sigma},c=e^{\sigma}\cr
\gamma=e^{\phi-\chi}\equiv{e^\phi}\eta\cr
\beta=e^{\chi-\phi}\partial\chi\equiv\partial\xi{e^{-\phi}}}}

and the BRST charge is
\eqn\grav{\eqalign{Q=Q_1+Q_2+Q_3\cr
Q_1=\oint{{dz}\over{2i\pi}}(cT-bc\partial{c}) \cr
Q_2=-{1\over2}\oint{{dz}\over{2i\pi}}(\gamma\psi_m\partial{X^m}
-q\partial\lambda)
\cr
Q_3=-{1\over4}\oint{{dz}\over{2i\pi}}b\gamma^2
}}

Then, in the limit $\mu_0\rightarrow{0}$
 the action (3) is particularly symmetric under the following global
space-time transformations (for complete list of symmetries e.g see ~{\selfc})

\eqn\grav{\eqalign{\delta{X^m}=\epsilon^{m}{\lbrace}\partial(e^\phi\lambda)
+2e^\phi\partial\lambda\rbrace\cr
\delta\lambda=-\epsilon^m{\lbrace}e^\phi\partial^2{X_m}+2\partial(
e^\phi\partial{X_m})
\cr
\delta\gamma=\epsilon^m{e^{2\phi-\chi}}
(\lambda\partial^2{X_m}-2\partial\lambda\partial{X_m})
\cr
\delta\beta=\delta{b}=\delta{c}=0}}

with the generator of (7) given by

\eqn\grav{\eqalign{T_m={1\over{\rho}}\oint{{dz}\over{2i\pi}}
e^\phi(\lambda\partial^2{X_m}-2\partial\lambda\partial{X_m})}}
where $\rho$ is some constant (which we shall relate to AdS
radius and cosmological constant, while relating $T_m$ to generator
of transvections).
This generator is not BRST-invariant and therefore the symmetry
transformations generated by (8) are incomplete
(similarly, the rotation generator
$T_{mn}=\oint{{dz}\over{2i\pi}}\psi_m\psi_n$ is not BRST invariant
and therefore only induces rotations for the $\psi$-fields but not for bosons).
To make both $T_m$ and $T_{mn}$ complete one has to 
restore their BRST invariance by adding ghost dependent correction terms.
These terms can be obtained by the homotopy
$K$-transformation described in ~{\selfc, \spinself}, 
which we shall briefly review below.
Let $Q$ be the BRST operator given by (6)
and let
\eqn\grav{\eqalign{
T=\oint{{dz}\over{2i\pi}}V(z)}}
be some global symmetry generator, incomplete (in the sense described above)
and not BRST invariant, satisfying
\eqn\grav{\eqalign{\lbrack{Q_{brst}},V(z)\rbrack=\partial{U}(z)+W(z)}}
and therefore
\eqn\lowen{\lbrack{Q_{brst}},T{\rbrack}=\oint{{dz}\over{2i\pi}}W(z)}
Introduce  the homotopy operator
\eqn\lowen{K(z)=-4c{e}^{2\chi-2\phi}(z)\equiv{\xi}\Gamma^{-1}(z)}
satisfying
\eqn\lowen{\lbrace{Q_{brst}},K(z)\rbrace=1}
In general, the homotopy operator has a non-singular OPE with $W$.
Suppose this OPE is given by
\eqn\lowen{K(z_1)W(z_2)\sim{(z_1-z_2)^N}Y(z_2)+O((z_1-z_2)^{N+1})}
where $N\geq{0}$ and $Y$ is some operator of dimension $N+1$.

Then the complete BRST-invariant symmetry generator ${{L}}$
can be obtained from the incomplete non-invariant symmetry generator
$T$ by the following homotopy transformation:
\eqn\grav{\eqalign{
T\rightarrow{{L}}(w)=K{\circ}T=T+{{(-1)^N}\over{N!}}
\oint{{dz}\over{2i\pi}}(z-w)^N:K\partial^N{W}:(z)
\cr
+{1\over{{N!}}}\oint{{dz}\over{2i\pi}}\partial_z^{N+1}{\lbrack}
(z-w)^N{K}(z)\rbrack{K}\lbrace{Q_{brst}},U\rbrace}}
where $w$ is some arbitrary point on the worldsheet
and $K{\circ}$ represents the transformation
(15) using the $K(z)$ operator (12).
It is straightforward to check the invariance
of ${{L}}$ by using some partial integration along with
the relation (13) as well as the obvious identity
\eqn\lowen{\lbrace{Q_{brst}},W(z)\rbrace=
-\partial(\lbrace{Q_{brst}},U(z)\rbrace)}
that follows directly from (10).
The homotopy transformed BRST-invariant ${{L}}$-generators
are then typically of the form 
\eqn\lowen{{{L}}(w)=\oint{{dz}\over{2i\pi}}(z-w)^N{\tilde{V}}_{N+1}(z)}
with the conformal dimension $N+1$ operator ${\tilde{V}}_{N+1}(z)$
in the integrand satisfying
\eqn\lowen{{\lbrack}Q_{brst},{\tilde{V}}_{N+1}(z)\rbrack
=\partial^{N+1}{\tilde{U}}_0(z)}
where ${\tilde{U}}_0$ is some operator of conformal dimension zero.
Although for $N>0$ the L-operator  depends on an arbitrary point 
on the worldsheet, such a dependence is irrelevant in correlation
functions since it can be shown ~{\selfc} that all the $w$-derivatives
of $L$ are BRST exact in small Hilbert space.
We shall refer to $L$ as homotopy image of $K$.
For our purposes, it will be also
 convenient to generalize the definitions (10)-(15) as follows.
Namely, we shall refer to operator $L=K_{\Upsilon}\circ{T}$ as
a $partial$ homotopy transform of $T$ based on $\Upsilon$, if
the operator $T=\oint{V}$ satisfies
$\lbrack{Q_1},{V}\rbrack=\partial(cU)+W$,
 $\Upsilon$ is some dimension 1 operator, the
OPE of $K$ and $\Upsilon$ is non-singular with the leading
order $N>0$  and $L$ is related to $K$ according to the transformation
(15) with $W$ replaced by $\Upsilon$, i.e.
\eqn\grav{\eqalign{
{{L}}(w)=K_\Upsilon{\circ}T=T+{{(-1)^N}\over{N!}}
\oint{{dz}\over{2i\pi}}(z-w)^N:K\partial^N{\Upsilon}:(z)
\cr
+{1\over{{N!}}}\oint{{dz}\over{2i\pi}}\partial_z^{N+1}{\lbrack}
(z-w)^N{K}(z)\rbrack{K}\lbrace{Q_{brst}},U\rbrace}}
Particularly, if $\lbrack{Q},T\rbrack=\oint\Upsilon$, 
the partial homotopy transform  obviously
coincides with the usual
homotopy transform (15).
In the following sections, we shall particularly use the partial
homotopy transforms in order to construct operators with necessary
on-shell conditions.

Let us now apply the above prescription  to the symmetry generators (1), (8).
The homotopy transformed full BRST-invariant rotation generator
is then given by

\eqn\grav{\eqalign{L_{mn}=\oint{{dz}\over{2i\pi}}
{\lbrack}\psi_m\psi_n+2ce^{\chi-\phi}\psi_{\lbrack{m}}\partial{X_{n\rbrack}}
-4\partial{c}ce^{2\phi-2\chi}\rbrack\cr
=-4\lbrace{Q},\xi{\Gamma^{-1}}\psi_m\psi_n\rbrace}}
Note that the generator (20) can be written as a BRST commutator in the
$large$ Hilbert space.
It is straightforward to check that the generator (20) induces 
(up to the terms, BRST exact in $small$ Hilbert space)
 Lorenz rotations for all the matter fields
(both $X$ and $\psi$).
Similarly,  the homotopy transformation of the generator (8)
gives full BRST-invariant
symmetry generator given by
\eqn\grav{\eqalign{L^m(w)=\oint{{dz}\over{2i\pi}}(z-w)^2
\lbrace{1\over2}P^{(2)}_{2\phi-2\chi-\sigma}{e^\phi}F^m_{5\over2}
-12\partial{c}ce^{2\chi-\phi}
F^m_{5\over2}\cr
+ce^\chi\lbrack
-{2\over3}\partial^3\psi^m\lambda+{4\over3}\partial^3\varphi\partial{X^m}
+2\partial^2\psi^m\partial\lambda
\cr
+P^{(1)}_{\phi-\chi}(-2\partial\varphi\partial^2{X^m}
+4\partial^2\varphi\partial{X^m}-2\partial^2\psi^m\lambda
+4\partial\psi^m\partial\lambda)
\cr
+P^{(2)}_{\phi-\chi}(2\partial\varphi\partial{X^m}+2\psi^m\partial\lambda-
2\partial\psi^m\lambda-q\partial^2{X^m})
+P^{(3)}_{\phi-\chi}(-{2\over3}\psi^m\lambda+{{4q}\over3}\partial{X^m})\rbrack
\rbrace
\cr
=-4{\lbrace}Q,
\oint{{dz}\over{2i\pi}}(z-w)^2
ce^{2\chi-\phi}F^m_{5\over2}(z)\rbrace}}
so the full vector symmetry generator is again the BRST commutator
in the large Hilbert space.
Here  $F^m_{5\over5}=\lambda\partial^2{X_m}-2\partial\lambda\partial{X_m}$
and the conformal weight $n$ polynomials
$P^{(n)}_{a\phi+b\chi+c\sigma}$ (where $a,b,c$ are some constants) are defined
according to
\eqn\lowen{P^{(n)}_{a\phi+b\chi+c\sigma}=e^{-a\phi(z)-b\chi(z)-c\sigma(z)}
{{d^n}\over{dz^n}}e^{a\phi(z)+b\chi(z)+c\sigma(z)}}
(with the product taken in algebraic rather than OPE sense).
The BRST-invariant symmetry generator can also be
be constructed at dual $-3$ picture (as well as the  pictures below;
 but not above minimal negative picture $-3$ 
at which it is annihilated by the picture
changing). At picture $-3$ the symmetry generator is given 
by

\eqn\lowen{L^m=\oint{{dz}\over{2i\pi}}e^{-3\phi}F^m_{5\over2}}

The symmetry generators (21), (23) at pictures $+1$ and $-3$
are related by the sequence of 
$Z$-transformations and the picture-changing
~{\selfc}
 according to 
$$L^m_{(+1)}=Z:\Gamma^2:{Z}:\Gamma^2:L^m_{(-3)}$$
where $\Gamma=\lbrace{Q},e^\chi\rbrace$ is the picture-changing operator
for the $\beta-\gamma$ system while
 $Z=b\delta(T)$ is the operator
of picture-changing for the $b-c$ system (particularly, it maps
unintegrated vertex operators to integrated). The manifest
integral form of $Z$ is given e.g. in ~{\spinself}.
 
With some effort, it can now be shown that the operators
$L^{mn}$ and $L_m$ realize the $AdS_{d}$ isometry algebra
(2) with the cosmological constant $\Lambda=-{1\over{\rho^2}}$

To demonstrate this, we start with the OPE of the the primary
fields $F^m_{5\over2}(z)$ (related to the matter ingredient of $L^m$)
Straightforward calculation gives
\eqn\grav{\eqalign{F^m_{5\over2}(z)F^n_{5\over2}(w)=
-{{6\eta^{mn}}\over{(z-w)^5}}
+{{14\partial\lambda\lambda\eta^{mn}(w)+8\partial{X^m}\partial{X^n}(w)}\over
{(z-w)^3}}
\cr
+
{{10\partial^2{X^m}\partial{X^n}(w)-2\partial{X^m}\partial^2{X^n}(w)
+7\eta^{mn}\partial^2\lambda\lambda(w)}\over{(z-w)^2}}
\cr
+{{6\partial^3{X^m}\partial{X^n}(w)-3\partial^2{X^m}\partial^2{X^n}(w)
+3\eta^{mn}\partial^3\lambda\lambda(w)+
2\eta^{mn}\partial^2\lambda\partial\lambda(w)}
\over{z-w}}
\cr
+(z-w)^0{\lbrack}
{7\over3}\partial^4{X^m}\partial{X^n}(w)
-2\partial^3{X^m}\partial^2{X^n}(w)
\cr
+{{11}\over{12}}\eta^{mn}\partial^4\lambda\lambda(w)
+{4\over3}\eta^{mn}\partial^3\lambda\partial\lambda(w)\rbrack
+:{F^m_{5\over2}}{F^m_{5\over2}}:(w)
\cr
+(z-w){\lbrack}
{{13}\over{60}}\eta^{mn}\partial^5\lambda\lambda(w)
-{1\over2}\eta^{mn}\partial^4\lambda\partial\lambda(w)
+
{2\over3}\partial^5{X^m}\partial{X^n}(w)
\cr
-{5\over6}2\partial^4{X^m}\partial^2{X^n}(w)
+:\partial{F^m_{5\over2}}{F^m_{5\over2}}:(w)
\rbrack\cr
+(z-w)^2
\lbrack
{1\over{24}}\eta^{mn}\partial^6\lambda\lambda(w)
-{2\over{15}}\eta^{mn}\partial^5\lambda\partial\lambda(w)
\cr
+{3\over{20}}
\partial^6{X^m}\partial{X^n}(w)
-{1\over4}
\partial^5{X^m}\partial^2{X^n}(w)+
{1\over2}:\partial^2{F^m_{5\over2}}{F^m_{5\over2}}:(w)
+...}}
Using this OPE it is straightforward to compute the commutator
${\lbrack}L^m,L^n{\rbrack}$. It is convenient to choose one of the
vector at picture $+1$ representation (21) and another at negative 
picture $-3$  representation (23). 
Because of the isomorphism between positive and negative
picture representations, ensured by the
appropriate $Z,\Gamma$ transformations (see below equation (23)),
the fiunal result will be picture-independent.
Then , using (21) and the BRST invariance
of $L^{m}$  at negative picture (23), we get

\eqn\grav{\eqalign{{\lbrack}L^m,L^n{\rbrack}=
{1\over{\rho^2}}\lbrace{Q},\lbrack\oint{{dz_1}\over{2i\pi}}(z_1-w)^2
{c}e^{2\phi-\chi}{F^m_{5\over2}}(z_1),\oint{{dz_2}\over{2i\pi}}
e^{-3\phi}{F^m_{5\over2}(z_2)}\rbrack\rbrace
\cr
=
\lbrace{Q,U(z_2)}\rbrace}}
where
\eqn\grav{\eqalign{
U(z_1)\equiv\oint{{dz_2}\over{2i\pi}}U_1(z_2)+
\oint{{dz_2}\over{2i\pi}}(z_1-z_2)
U_2(z_2)+\oint{{dz_2}\over{2i\pi}}(z_1-z_2)^2U_3(z_2)
\cr 
=\oint{{dz_2}\over{2i\pi}}
ce^{2\chi-4\phi}{\lbrack}
{7\over3}
\partial^4{X^{{\lbrack}m}}\partial{X^{n\rbrack}}
-2\partial^3{X^{{\lbrack}m}}\partial^2{X^{n\rbrack}}
+6P^{(1)}_{2\chi-\phi+\sigma}\partial^3{X^{{\lbrack}m}}\partial{X^{n\rbrack}}
\cr
+4P^{(2)}_{2\chi-\phi+\sigma}\partial^3{X^{{\lbrack}m}}\partial{X^{n\rbrack}}
+:{F^m_{5\over2}}{F^m_{5\over2}}:\rbrack
\cr
+
\oint{{dz_2}\over{2i\pi}}(z_1-z_2)ce^{2\chi-4\phi}
\lbrack
{4\over3}P^{(3)}_{2\chi-\phi+\sigma}\partial^2{X^{{\lbrack}m}}\partial{X^{n\rbrack}}
+3P^{(2)}_{2\chi-\phi+\sigma}\partial^3{X^{{\lbrack}m}}\partial{X^{n\rbrack}}
\cr
+P^{(1)}_{2\chi-\phi+\sigma}({7\over3}
\partial^4{X^{{\lbrack}m}}\partial{X^{n\rbrack}}
-2\partial^3{X^{{\lbrack}m}}\partial^2{X^{n\rbrack}})
\cr
+{2\over3}
\partial^5{X^{{\lbrack}m}}\partial{X^{n\rbrack}}
-{5\over6}\partial^4{X^{{\lbrack}m}}\partial^2{X^{n\rbrack}}
\rbrack\cr
+
\oint{{dz_2}\over{2i\pi}}(z_1-z_2)^2{ce^{2\chi-4\phi}}
\lbrack
{1\over3}P^{(4)}_{2\chi-\phi+\sigma}\partial^2{X^{{\lbrack}m}}\partial{X^{n\rbrack}}
+
P^{(3)}_{2\chi-\phi+\sigma}\partial^3{X^{{\lbrack}m}}\partial{X^{n\rbrack}}
+
\cr
P^{(2)}_{2\chi-\phi+\sigma}({7\over6}
\partial^4{X^{{\lbrack}m}}\partial{X^{n\rbrack}}
-\partial^3{X^{{\lbrack}m}}\partial^2{X^{n\rbrack}})
\cr
+
P^{(1)}_{2\chi-\phi+\sigma}({2\over3}
\partial^5{X^{{\lbrack}m}}\partial{X^{n\rbrack}}
-{5\over6}\partial^4{X^{{\lbrack}m}}\partial^2{X^{n\rbrack}})
+{3\over{20}}\partial^6{X^{{\lbrack}m}}\partial^2{X^{n\rbrack}}
-{1\over4}\partial^5{X^{{\lbrack}m}}\partial^2{X^{n\rbrack}}\rbrack}}

where, for convenience, we  split
the overall integral into 3 parts, with the integrands proportional
to $U_1(z_2)$, $(z_1-z_2)U_2(z_2)$ and $(z_1-z_2)^2U(z_2)$ accordingly.

To relate the right hand side of the commutator
(26) 
to the rotation generator (20) one has to perform double picture changing
transform of $\lbrace{Q,U(z_1)}\rbrace$ in order to bring it to picture zero.
We shall demonstrate the procedure explicitly 
for the $U_1$ integral, with  the other two integrals treated similarly. 
For that, we first of all need a manifest expression for the commutator
of the BRST charge with the $U_1(z_1)$ operator (26). Straightforward
calculation gives:
\eqn\grav{\eqalign{
\lbrace{Q,U_1(z)}\rbrace
=-2\partial{c}c{e^{2\chi-4\phi}}(P^{(2)}_{2\chi-\phi+2\sigma}
+P^{(2)}_{2\chi-\phi+\sigma})\partial^2{X^{{\lbrack}m}}\partial{X^{n\rbrack}}
\cr
+9
\partial^2{c}c{e^{2\chi-4\phi}}P^{(1)}_{2\chi-\phi+\sigma}\partial^2{X^{{\lbrack}m}}
\partial{X^{n\rbrack}}
\cr
-{3\over2}
\partial^2{c}c{e^{2\chi-4\phi}}\partial^3{X^{{\lbrack}m}}\partial{X^{n\rbrack}}
+
\partial{c}c{e^{2\chi-4\phi}}
({7\over3}
\partial^4{X^{{\lbrack}m}}\partial{X^{n\rbrack}}
-2\partial^3{X^{{\lbrack}m}}\partial{X^{n\rbrack}}
)
\cr
+{{34}\over3}
\partial^3{c}c{e^{2\chi-4\phi}}\partial^2{X^{{\lbrack}m}}\partial{X^{n\rbrack}}
+12\partial^2{c}c{e^{2\chi-4\phi}}\partial^3{X^{{\lbrack}m}}\partial{X^{n\rbrack}}
\cr
+4ce^{\chi-3\phi}
\lbrack
\partial^2{\psi^{{\lbrack}m}}\partial{X^{n\rbrack}}
+\partial{\psi^{{\lbrack}m}}\partial^2{X^{n\rbrack}}
-P^{(1)}_{\phi-\chi}(
{\psi^{{\lbrack}m}}\partial^2{X^{n\rbrack}}
-2\partial{\psi^{{\lbrack}m}}\partial{X^{n\rbrack}}
)
\cr
+
{\psi^{{\lbrack}m}}\partial{X^{n\rbrack}}
(P^{(2)}_{\phi-\chi}+P^{(2)}_{2\chi-\phi+\sigma})
\rbrack
}}
The next step is to perform the normal ordering of
the integrand of this expression with $\xi=e^\chi$ around the midpoint.
We get
\eqn\grav{\eqalign{
:\xi\lbrace{Q,U_1(z)}\rbrace:
=
-8\partial{c}c{e^{3\chi-4\phi}}\partial^2{X^{{\lbrack}m}}\partial{X^{n\rbrack}}
+4ce^{2\chi-3\phi}
\lbrack
{\psi^{{\lbrack}m}}\partial^2{X^{n\rbrack}}
\cr
+4\partial{\psi^{{\lbrack}m}}\partial{X^{n\rbrack}}
-2{\psi^{{\lbrack}m}}\partial{X^{n\rbrack}}P^{(1)}_{3\phi-\chi-2\sigma}
\rbrack}}
The next step is to perform the commutation of this expression with $Q$
which, by definition, gives us
$\lbrace{Q,U(z_1)}\rbrace$ at picture $-1$. Straightforward calculation
gives:
\eqn\grav{\eqalign{
\lbrace{Q},
\xi\lbrace{Q,U_1(z)}\rbrace\rbrace
=ce^{\chi-2\phi}(4P^{(1)}_{\phi+\chi-2\sigma}\psi^m\psi^n+2\psi^{\lbrack{m}}
\partial\psi^{n\rbrack})}}
Next, the normal ordering of this expression with $\xi$
around the midpoint gives
\eqn\grav{\eqalign{
:\xi\lbrace{Q},
\xi\lbrace{Q,U_1(z)}\rbrace\rbrace:
=4ce^{2\chi-2\phi}\psi^m\psi^n}}
Finally, the commutator of this expression with Q by definition gives
us $U_1$ at picture zero:

\eqn\grav{\eqalign{U_1^{(0)}(z)\equiv
\lbrace{Q},\xi\lbrace{Q},
\xi\lbrace{Q,U_1(z)}\rbrace\rbrace\rbrace
=4\lbrace{Q},e^{2\chi-2\phi}\psi^m\psi^n\rbrace}}

which, according to (20) is nothing but the integrand
of the full rotation generator with the inverse sign.
The picture transform of $U_2$ and $U_3$ in the remaining
terms of (20) is performed similarly. Applying picture changing transformation
twice and integrating out total derivatives we find the contributions from the
 second  and the third integrals cancel each other and the commutator
remains unchanged. This concludes the proof that
 the commutator of two operators
${\lbrack}L^m,L^n\rbrack=-{1\over{\rho^2}}L^{mn}$
reproduces the commutation of two transvections in the $AdS$ isometry algebra
(2). 

The remaining commutators of (2) are computed similarly.
Note the highly nontrivial appearance of the minus sign on the right
hand side of the commutator as a result of the ghost structure of the 
operators, that indicates that the effective cosmological constant
in the symmetry algebra is negative, so the  effective geometry
generated is of the $AdS$  type.

The combination of operators $(L^m,L^{mn})$ (20), (21) that we considered
so far, 
is not the only possible realization of the AdS symmetry algebra (2) 
in RNS theory.
In particular, it is easy to check that so(d-1,2)
isometry of the AdS space is realized by  $S^m,L^{mn}$
where 
\eqn\lowen{
L^{mn}=K\circ{T^{mn}}\equiv{K}\circ\oint{{dz}\over{2i\pi}}\psi^m\psi^n}
is the same full rotation operator (20) 
(where the $K\circ$ represents the homotopy transformation
to ensure the BRST-invariance)
while $S^{m}$ is the homotopy transformation of the operator
$\oint{{dz}\over{2i\pi}}\lambda\psi^m$, representing
the rotation in the Liouville-matter plane:
\eqn\grav{\eqalign{
S^m=K\circ{\rho^{-1}}\oint{{dz}\over{2i\pi}}\lambda\psi^m
\cr
={{\rho^{-1}}}\oint{{dz}\over{2i\pi}}\lbrack
\lambda\psi^m+2ce^{\chi-\phi}(\partial\varphi\psi^m-\partial{X^m}\lambda
-qP^{(1)}_{\phi-\chi}\psi^m)
-4\partial{c}c{e^{2\chi-2\phi}}\lambda\psi^m\rbrack
\cr
=-4{\lbrace}Q,{\rho^{-1}}
\oint{{dz}\over{2i\pi}}ce^{2\chi-2\phi}\lambda\psi^m\rbrace}}
Again, using (33) it and the procedure identical to
the one explained above, it is straightforward to show
that $S^m$ satisfy the commutation relation for transvections:
$\lbrack{S^m,S^n}{\rbrack}=-{{{L^{mn}}\over{\rho^2}}}$
and the rest of $so(d-1,2)$ relations (2) with $L^{mn}$.
In order to construct vertex operators for spin connection
in $AdS$ space, we will actually need the realization using the
linear combination of the transvections (21),(33), given by
\eqn\lowen{P^{m}={1\over{{\sqrt{2}}}}(L^m+S^m)}
One can show that these generators realize the transvections
on $AdS$ space provided that the two-form (20) is shifted
according to
\eqn\grav{\eqalign{L^{mn}\rightarrow{P^{mn}}(w)=L^{mn}+
K\circ\oint{{dz}\over{2i\pi}}e^{-3\phi}(\psi^{\lbrack{m}}\partial^2{X^{n\rbrack}}
-2\partial\psi^{\lbrack{m}}\partial{X^{n\rbrack}})\cr
+qK\circ
\oint{{dz}\over{2i\pi}}c{e^{\chi-\phi}}\lambda\psi^m\psi^n}}
where $K$ is again the homotopy transformation.
The new term, proportional to the
 background charge, appears as a result of the Liouville terms
of $Q$ entering the game of picture changing.
The AdS isometry algebra (2)  is then realized by 
the combination of $P^{m}$ and $P^{mn}$.
In the next section, we will use these $AdS$ isometry generators as building
blocks to construct vertex operators for $AdS$ frame fields in
closed string theory.

\centerline{\bf 3. Vertex operators for frame fields and on-shell conditions }

In this section we construct vertex operators  for connection
gauge fields for MMSW gravity with cosmological constant, using the
generators (20),(21), (33)-(35) 
and study their BRST properties. We find that the
BRST invariance constraints leads to linearized equations of motion
for MMSW gravity around the AdS vacuum, while the nontriviality constraints
entail the gauge transformations for the frame fields.
To start with, let us recall the basic facts about MMSW formulation of gravity.
In the frame-like approach, the description
of the dynamics of the theory in terms of
metric tensor $g_{mn}$ is replaced by introducing two dynamical fields
- the frame field $e_m^a$ and the connection gauge field
$\omega_m^{ab}$ with the indices $a$ and $b$ living in the tangent space.
Using these fields one constructs one-forms
$e^a=e^{a}_mdx^m$ and $\omega^{ab}=\omega^{ab}_mdx^m$ and unifies them
into a single one-form
$\omega=e^a{T_a}+{1\over2}\omega^{am}T_{ab}$ where $T_a$ and $T_{ab}$
are isometry generators of $AdS_d$.
The curvature is then the two-form defined according
to
\eqn\lowen{R^{ab}=d\omega^{ab}+\omega^a_c\wedge\omega^{cb}-\rho^{-2}e^a\wedge
e^{b}}
while the two-form of torsion is given by
\eqn\lowen{T^a=de^a+\omega^a_c\wedge{e^c}}.
It is convenient to unify the connection and the frame fields
into a single $o(d-1,2)$ gauge field
\eqn\lowen{\omega^{AB}\equiv(\omega^{ab},\omega^{a{\hat{d}}})}
where by definition $\omega^{a{\hat{d}}}=\rho^{-1}e^a$
and the $o(d-1,2)$ index is split in the $(d,1)$ way as
$A\equiv(a,{\hat{d}})$.
The curvature and the torsion are then unified
into a single tensor $R^{AB}=(R^{ab},R^{a{\hat{d}}})$
with $R^{a{\hat{d}}}=\rho^{-1}T^a$.
The $AdS_d$ geometry is then the solution of the 
vacuum equations 
\eqn\lowen{R^{AB}(\omega)=0}
which combine the constant curvature and the zero torsion constraints.
The gauge symmetry transformations for
the 1-form  $\omega^{AB}=\omega_m^{AB}dx^m$
are given by 
\eqn\grav{\eqalign{\delta^{gauge}\omega_m^{AB}=D_m\rho^{AB}
\cr
\delta^{diff}\omega_m^{AB}=\partial_m\epsilon^n\omega_n^{AB}
+\epsilon^{n}\partial_n\omega_m^{AB}}}
where $\rho^{AB}$ and $\epsilon^m$ are the parameters of the gauge 
and diffeomorphism transformations accordingly.

Our goal now will be to construct a sigma-model based on vertex operators
for connection and frame gauge fields, which beta-functions reproduce (39).
Since AdS geometry would appear as the vacuum solution of (39),
the low-energy limit  of the string
theory sigma-model we are looking for would describe the 
MMSW gravity on anti-de Sitter space in the frame-like formulation.
Just as a standard graviton operator (describing fluctuations of metric
around flat vacuum) is given by the structure bilinear in translation 
operators (multiplied by $e^{ipX}$), we shall look for vertex operators
for the MMSW gauge fields as closed string bilinears based on generators
(20),(21),(33)-(35) realizing the $AdS_d$ isometry  (2).
The operator that we propose
is given by
\eqn\grav{\eqalign{
G(p)=e^a_m(p)F_a{\bar{L}}^m+
+\omega^{ab}_m(p)
(F^m_{b}{\bar{L}}_a-{1\over2}F_{ab}{\bar{L}}^m)+c.c.}}
where
\eqn\grav{\eqalign{
F_m=-2K_{U_1}\circ\int{dz}\lambda\psi_me^{ipX}(z)\cr
U_1=\lambda\psi_me^{ipX}+{i\over2}\gamma\lambda
(({\vec{p}}{\vec{\psi}})\psi_m-p_mP^{(1)}_{\phi-\chi})e^{ipX}}}
or manifestly
\eqn\grav{\eqalign{
F_m=-2\int{dz}{\lbrace}\lambda\psi_m(1-4\partial{c}ce^{2\chi-2\phi})+
\cr
2ce^{\chi-\phi}
(\lambda\partial{X}_m-\partial\varphi\psi_m+q\psi_mP^{(1)}_{\phi-\chi}
-{i\over2}(({\vec{p}}{\vec{\psi}})\psi_m-p_mP^{(1)}_{\phi-\chi}))\rbrace{e^{ipX}}(z)}}
Next,
\eqn\grav{\eqalign{{\bar{L}}^a=\int{d{\bar{z}}}e^{-3{\bar\phi}}
\lbrace{\bar\lambda}\bar\partial^2{X^a}-2\bar\partial\bar\lambda\bar\partial
{X^a}
\cr+
ip^a({1\over2}\bar\partial^2\bar\lambda+{1\over{q}}
\bar\partial\bar\varphi\bar\partial\bar\lambda
-{1\over2}\bar\lambda(\bar\partial\bar\varphi)^2+
(1+3q^2)\bar\lambda(3\bar\partial\bar\psi_b\bar\psi^b-{1\over{2q}}
\bar\partial^2\bar\varphi))\rbrace{e^{ipX}}}}
(similarly for its holomorphic counterpart $L^a$)
and
\eqn\grav{\eqalign{
F_{ma}=F_{ma}^{(1)}+F_{ma}^{(2)}+F_{ma}^{(3)}}}
where
\eqn\grav{\eqalign{
F_{ma}^{(1)}=-4qK_{U_2}\circ\int{dz}ce^{\chi-\phi}\lambda\psi_m\psi_a\cr
U_2=\lbrack{Q-Q_3},ce^{\chi-\phi}\lambda\psi_m\psi_a{e^{ipX}}\rbrack
-{i\over2}c\lambda(({\vec{p}}{\vec{\psi}})\psi_a\psi_m-p_m\psi_aP^{(1)}_{\phi-\chi})
e^{ipX}(z)}}
\eqn\grav{\eqalign{
F_{ma}^{(2)}=K\circ\int{dz}\psi_m\psi_a{e^{ipX}}=-4\lbrace{Q},\int{dz}
ce^{2\chi-2\phi}{e^{ipX}}\psi_m\psi_a(z)\rbrace}}
and
\eqn\grav{\eqalign{
F_{ma}^{(3)}=\int{dz}e^{-3\phi}(\psi_{\lbrack{m}}\partial^2{X}_{a\rbrack}
-2\partial\psi_{\lbrack{m}}\partial{X}_{a\rbrack})e^{ipX}(z)}}
In the limit  of zero momentum the holomorphic and the antiholomorphic
components of the operator (41) correspond to AdS isometry generators
(20), (21), (33)-(35) in different realizations, described above.
More precisely,
while the antiholomorphic part of (41)
is based on the L- operators (20),(21)
related to  the $L$-realization  of the symmetry algebra,
 the holomorphic part of (41)
 involves the F-operators (such as $F^{a}$ and $F^{ab}$)
which, although different from the operators of the $P$-representation,
become related to those after one imposes the on-shell constraints
on the space-time fields (see below).

We start with analyzing the BRST invariance constraints on the operator (41).
The BRST commutators are given by:

\eqn\grav{\eqalign{
\lbrack{\bar{Q}},G(p)\rbrack=0\cr
\lbrack{\bar{Q}},G(p)\rbrack
=ie_m^b(p){\bar{L}}_b\int{dz}\gamma\lambda
(({\vec{p}}{\vec{\psi}})\psi_m-P^{(1)}_{\phi-\chi}{p_m})e^{ipX}(z)
\cr
+\omega_m^{ab}{\bar{L}}_b\int{dz}\lbrace
\gamma\lambda\psi_a\psi_m+2ic\lambda
({\vec{p}}{\vec{\psi}})\psi_a\psi_m-P^{(1)}_{\phi-\chi}p_m\psi_a
\rbrace{e^{ipX}}(z)}}
The BRST invariance therefore imposes the following constraints
on vielbein and connection fields:
\eqn\grav{\eqalign{
p^{\lbrack{n}}e^b_{m\rbrack}(p)-\omega^{b{\lbrack}n}_{m\rbrack}(p)=0\cr
p_{\lbrack{n}}\omega_{m\rbrack}^{ab}(p)=0
\cr
p^m{e_m^b}(p)=0
\cr
p^m\omega_m^{ab}(p)=0}}
The first two constraints represent the linearized
equations $R^{AB}=0$ (the first one being the zero torsion constraint
$T^a=R^{a{\hat{d}}}=0$
whle the second reproducing vanishing Lorenz curvature $R^{ab}=0$).
The last two constraints  represent the gauge fixing conditions
related to the diffeomorphism symmetries (40).
The fact that the BRST invariance leads to space-time equations in
a certain gauge is not surprising if we recall that similar constraints
on a standard vertex operator of a photon also lead to Maxwell's equations in 
the Lorenz gauge.
Provided that the constraints (50) are satisfied the vertex operator $G(p)$
can be written as a BRST commutator in the large Hilbert space
plus terms that are manifestly in the small Hilbert space,
according to
\eqn\grav{\eqalign{
G(p)=\lbrace{Q},W(p)\rbrace+{1\over{q}}\omega_m^{ab}\int{dz}
e^{-3\phi}(\psi^{\lbrack{m}}\partial^2{X}_{a\rbrack}
-2\partial\psi_{\lbrack{m}}\partial^{X}_{a\rbrack})e^{ipX}(z)
{\bar{L}}_b+c.c.
\cr
W(p)=8e^a_m(p){\bar{L}}_a\int{dz}
c\partial\xi\xi{e^{-2\phi}}\lambda\psi^me^{ipX}
\cr
+\omega_m^{ab}{\bar{L}}_b{\lbrack}-{4\over{q}}\int{dz}
c\partial\xi\xi{e^{-2\phi}}\psi_a\psi^m{e^{ipX}}
\cr
+4\int{dz}(z-w)\partial{c}c\partial^2\xi\partial\xi\xi{e^{-3\phi}}
\lambda\psi_a\psi^m{e^{ipX}}\rbrack}}
This particularly implies that , modulo gauge transformations,
the vertex operator $G(p)$ is the element of the $small$ Hilbert space.
Let us now turn to the question 
 of BRST nontriviality and related gauge symmetries
(40).
The linearized gauge symmetry transformations (40)
are given by
\eqn\grav{\eqalign{\delta{e_m^a}=\partial_m\rho^{a}+\rho_m^a\cr
\delta\omega_m^{ab}=\partial_m\rho^{ab}+\rho^{\lbrack{a}}\delta^{b\rbrack}_m}}
where we write $\rho^{AB}=(\rho^{ab},\rho^{a{\hat{d}}})=(\rho^{ab},\rho^a)$
The variation of $G(p)$ under (52) in the momentum space is 
\eqn\grav{\eqalign{\delta{G(p)}=
p^mF_m{\bar{L}}_a\rho^a+ p^mF_{ma}{\bar{L}}_b\rho^{ab}}}
The  two terms of the variation (53) are BRST exact in the $small$ 
Hilbert space (and therefore are irrelevant in correlators) 
since
\eqn\grav{\eqalign{
p^mF_m=\lbrace{Q}, :\Gamma:(w)\lbrack{Q},\xi{A}\rbrack\rbrace
\cr
A=\int{dz}e^{\chi-3\phi}\partial\chi(({\vec{p}}{\vec{\partial{X}}})\lambda
-({\vec{p}}{\vec\psi})\partial\varphi+({\vec{p}}{\vec\psi})
P^{(1)}_{\phi-(1+q)\chi}
)e^{ipX}}}

and

\eqn\grav{\eqalign{p^mF_{ma}^{(1)}=
4q\lbrack{Q},\Gamma(w)\int{dz}ce^{-3\phi}\partial\xi\partial^2\xi
\lambda\psi_a({\vec{p}}{\vec{\psi}})e^{ipX}\rbrack\cr
p^mF_{ma}^{(2)}=\lbrace{Q},:\Gamma:(w)
\int{dz}\partial\xi{e^{-3\phi}}(({\vec{p}}{\vec\psi})\partial{X}_a
-({\vec{p}}{\vec{\partial{X}}})\psi_a)e^{ipX}\rbrace\cr
p^mF_{ma}^{(3)}=\lbrace{Q},\lbrack{K}\circ\int{dz}\lambda\psi_a{e^{ipX}},B\rbrack
\rbrace\cr
B=\int{dz}\partial\xi{e^{-4\phi}}\lbrack
\lambda(\partial{\vec\psi}\partial^2{\vec{X}})
-2\partial\lambda(({\vec\psi}\partial^2{\vec{X}})-
2(\partial{\vec\psi}\partial{\vec{X}}))\rbrack
}}

Therefore gauge transformations of $e$ and $\omega$ shift 
$G(p)$ by terms not contributing to correlators.
This concludes the BRST analysis of the vertex operator for
vielbein and connection fields in the frame-like description of
MMSW gravity. In the next section we shall investigate 
the conformal beta-function of $G(p)$ in the sigma-model, showing
that it reproduces the equations of motion of  MMSW 
gravity with negative cosmological constant in the low energy limit.

\centerline{\bf 4. $\beta$-Function of $G(p)$ and AdS Gravity}

The leading order contribution to the beta-function of
the $G(p)$ operator
(giving the equations of motion for $e$ and $\omega$ in the low 
energy limit of string theory) is determined by the structure
constants stemming from three-point correlators on the worldsheet.
Computing these structure constants will be our goal in this section.
Manifest expressions for the operators (41)-(48) look quite
lengthy and complicated.
The computations, however, can be simplified significantly 
due to important property of the homotopy transformations (15):
That is, consider two operators
$V_1(z)$ and $V_2(w)$ (of dimension 1)
 that are, in general, not BRST-invariant and
are the elements of the small space (i.e. independent on  zero mode
 of $\xi$). Suppose their operator  products with the
homotopy operator $K$ are nonsingular while their full OPE between themselves
is given by
\eqn\grav{\eqalign{
V_1(p_1;z)V_2(p_2;w)=\sum_{k=-\infty}^{\infty}(z-w)^kC_k(p_1,p_2)V_k
(p_1+p_2;{{z+w}\over2})}}
where $C^k$ are the OPE coefficients and $V_k$ are some operators.
Then the operator product of their BRST-invariant homotopy
transforms is given by

\eqn\grav{\eqalign{
K_U\circ{V_1(p_1;z)}K_U\circ{V_2(p_2;w)}=\sum_{k=-\infty}^{\infty}(z-w)^kD_k(p_1,p_2)
K_U{\circ}W_k
(p_1+p_2;{{z+w}\over2})}}
with the coefficients $D_k$ and operators $W_k$ defined as follows.

Let $K_U\circ{V_1}$ and $K_u\circ{V_2}$
are the transforms of $V_1$ and $V_2$  that are BRST-invariant
(given the appropriate on-shell conditions on space-time fields)
Then they can be represented as BRST commutators in the large space:
$K_U\circ{V_1}=\lbrace{Q},KW_1\rbrace$ and
$K_U\circ{V_2}=\lbrace{Q},KW_2\rbrace$ where $W_1$ and 
$W_2$ are (generally) some new operators in the small space
( in many important cases $W_1$ and $W_2$ may actually coincide with
$V_1$ and $V_2$)
Let the full OPE of $W_1$ and $W_2$ be given by
\eqn\grav{\eqalign{
W_1(p_1;z)W_2(p_2;w)=\sum_{k=-\infty}^{\infty}(z-w)^kD_k(p_1,p_2)W_k
(p_1+p_2;{{z+w}\over2})}}
with certain operators and coefficients $W_k$ and $D_k$.
Then the OPE of the homotopy transforms
$K_U\circ{V_1}(z)$ and
$K_U\circ{V_1}(w)$ is given by the formula (57).
Indeed,

\eqn\grav{\eqalign{
K_U\circ{V_1}(z)K_U\circ{V_2}(w)
 =
\lbrace{Q},KW_1\rbrace(z)\lbrace{Q},LW_2\rbrace(w)
=\lbrace{Q},V-K\lbrack{Q,V}\rbrack(z):LV:(w)
\cr
=\lbrace{Q},\sum_{k=-\infty}^{\infty}K(w)D_k(p_1,p_2)W_k
(p_1+p_2;{{z+w}\over2})\rbrace
-\lbrace{Q},K\lbrack{Q},W_1\rbrack(z)KW_2(w)\rbrace}}
where we used the BRST invariance of
$\lbrace{Q},KW_1\rbrace$ and the OPE (58) of $W_1$ and $W_2$.
The OPE (59) is then given by
\eqn\grav{\eqalign{
K_U\circ{V_1}(z)K_U\circ{V_2}(w)
 =K_U\circ(W_1(z)W_2(w))-
\lbrace{Q},L\lbrack{Q},V\rbrack(z)LV(w)\rbrace}}
The first term in this OPE coincides with the right hand side of (57).
The second term is the BRST commutator in the small Hilbert space.
Indeed, if the OPEs of $K$ with $W_1$ and $W_2$ are nonsingular, one
can cast the second term in (60) as
\eqn\grav{\eqalign{
\lbrace{Q},K\lbrack{Q},W_1\rbrack(z)KW_2(w)\rbrace=
\lbrace{Q,C(z,w)}\rbrace\cr
C(z,w)=\sum_{m=0}^\infty(z-w)^m{W_1(z)}\lbrack{Q},\partial^m{L}LW_2\rbrack(w)}}
Since $C(z,w)$ is the product 
of $W_1$ (operator in the small Hilbert space) and the BRST commutator in
the large Hilbert space, it is the element of the small Hilbert space.
This concludes the proof of the formula (57), up to BRST exact
terms in the  small space, irrelevant for the beta-function.
The relation (57) is remarkably useful, since it allows us
to replace the computation of the products of homotopy-transformed
operators (which manifest expressions are cumbersome and complicated) 
with the products of operators which structure is far simpler.
The sigma-model we consider is given by:
\eqn\lowen{Z(e,\omega)=\int{D{\lbrack}X,\psi,\bar\psi,ghosts\rbrack}
e^{-S_{RNS}+\int{d^d}pG(p)}}
The leading order contributions to the $\beta$-function are given 
by terms quadratic in $G(p)$ and are proportional to $e^2$, $\omega^2$
and $e\omega$. Consider the contribution proportional to $e^2$ first.
It is given by
\eqn\grav{\eqalign{
{1\over2}\int_p\int_q{e_m^a}(p)e_n^b(q)(F^m{\bar{L}}_a(p)F^n{\bar{L}}_b(q)
+c.c.)
\cr
=
\int_p\int_q
{e_m^a}(p)e_n^b(q)\lbrace
(L^m+{K}\circ\int{dz}\lambda\psi^m){\bar{L}}_a
(L^n+{K}\circ\int{dw}\lambda\psi^n){\bar{L}}_b
+c.c.\rbrace
\cr
=-{1\over{\rho^2}}\int_{p}\int_q\int{{d^2\xi_1}\over{|\xi_1|^2}}
e_m^a(p)e_n^b(q)(F^{mn}{\bar{L}}_{ab}(p+q)+c.c.)\cr
=-{1\over{2\rho^2}}log\Lambda\int_p\int_q{e(p)\wedge{e}(q)}
(F{\bar{L}}(p+q)+c.c.)
 }}.
where $\xi_1=z_1-z_2$, $\Lambda$ is worldsheet cutoff and
we used (25)-(35) and the homotopy OPE property (57), as well as the fact
that the operator in front of the exponent in the expression
for $L^a$ (similarly for ${\bar{L}}^a$ has no OPE singularities
with $e^{ipX}$ or $e^{iqX}$ , up to BRST-exact terms.
One can easily recognize this logarithmic divergence contributing
the cosmological term to the low energy effective equations of motion.
Similarly, the
  term quadratic in $\omega$ 
contributes to the beta-function as
\eqn\grav{\eqalign{
{1\over2}\int_p\int_q\omega_m^{ab}(p)\omega_n^{cd}(q)
(F^m_a{\bar{L}}_b-{1\over2}F_{ab}{\bar{L}}^m+c.c.)(p)
(F^n_c{\bar{L}}_d-{1\over2}F_{cd}{\bar{L}}^n+c.c.)(q)
\cr
=\int_{p}\int_q\int{{d^2\xi_1}\over{|\xi_1|^2}}
{\lbrace}\omega_{\lbrack{m}}^{ab}(p)\omega_{n\rbrack}^{ad}(q)
(F^{mn}{\bar{L}}_{bd}+c.c.)(p+q)\rbrace
\cr
=
log\Lambda
\int_p\int_q
\lbrack{\omega(p)\wedge\omega(q)({F}{\bar{L}}+c.c.)(p+q)}\rbrack
}}
Thus the right-hand side  (64) accounts for 
$\omega\wedge\omega$ contribution to the $\beta$-function.

Also the divergence due to cross-terms proportional to
${\sim}e\omega$ vanishes provided
that the zero torsion constraint (50) is satisfied.

Altogether (50), (63), and (64)   imply the vanishing of the
conformal beta-function for
 the model (62) leads to the low-energy effective equations of motion:
\eqn\grav{\eqalign{R^{ab}=d\omega^{ab}+(\omega\wedge\omega)^{ab}-{1\over{\rho^2}}
e^a\wedge{e^b}=0}}
and
\eqn\lowen{
de^a+\omega^{ab}\wedge{e^b}=0}

which describe the AdS gravity in MMSW formalism.
The cosmological term with $\Lambda=-{1\over{\rho^2}}$ originates from
the transvection symmetry generators that serve as building blocks
for the vertex operators.
Thus the leading order contribution to 
the $\beta$-function in the sigma-model model (62)
describes the $AdS$ vacuum solution of the
MMSW gravity with negative cosmological constant.
As we only considered the lowest order 
contributions the beta-function (64), we only recovered the vacuum solution
with no fluctuations. 
The important next step will be to consider the fluctuations
of spin 2 and higher around the AdS vacuum. For that, one has to extend
(62) by adding terms with vertex operators, describing the higher spin
fluctuations in the frame-like approach, with some of these operators
constructed in ~{\selft} (see also (67) in the concluding Discussion section).
To describe the fluctuations of spins 2 and higher,
 around the $AdS$ vacuum, higher order corrections to the conformal
$\beta$-function of (62) and (67) need to be computed.
This calculation is currently in progress and we hope to present it soon
in our future work.

\centerline{\bf 5. Discussion. Higher Spin Dynamics on AdS and String Theory  }

The sigma-model considered in this work is constructed
to set up a framework for a string theory description of higher spin 
dynamics on AdS in Vasiliev's frame-like approach.
The basic idea is that  the dynamics of Vasiliev's frame-like fields
and generalized connections on AdS can be obtained from 
correlators of vertex operators for these gauge fields  in the presence
of the background $G(p)$-field constructed  in this work, which
effectively generates cosmological constanty and curves space-time 
from flat to AdS background.
That is, the generating functional for higher spin frame fields
$E^{a_1...a_1}$ and connections $\Omega^{a_1...a_n}$ should be
\eqn\lowen{
Z(E,\Omega,e,\omega)=\int{D}(X,\psi,\bar\psi,ghosts)e^{-S_{RNS}+G(p)
+E^{a_1...a_n}U_{a_1...a_n}+\Omega^{a_1...a_n}W_{a_1...a_n}}}
where $U$ and $W$ are the appropriate vertex operators for the 
higher spin gauge fields (we shall use capital letters for
higher spin connections and frame fields
to distinguish them from those in the theory of gravity)
The Vasiliev's unfolded equations for 
higher spin fields on $AdS$ space should then follow from
the worldsheet $beta$-function equations for the sigma-model (67).
The work in this direction is currently in progress. At this point
we have been able to investigate the model (67) for the $s=3$ case
in three dimensions, leading to higher spin dynamics on $AdS_3$.
Namely, the dynamic  gauge fields in the $s=3$ case are
given  by$E^{ab}_m$ and $\Omega^{ab}_m$ (which are spin 3 generalizations of
 frame and connection fields of MMSW theory) while 
the usual connection gauge field $\omega_m^{ab}$ 
can be dualized in $d=3$ as $\omega_{a}=\epsilon_{abc}\omega^{bc}$
The spin 3 vertex operator for  $E^{ab}$ is given by

\eqn\grav{\eqalign{U(p)=E^{ab}_m(p)\int{dz}{e^{-3\phi}}
\psi^m\partial{X_a}\partial{X_a}e^{ipX}(z)}}
or, in the positive picture representation,
\eqn\grav{\eqalign{U(p)=
E^{ab}_m(p)K\circ\int{dz}{e^{\phi}}
\psi^m\partial{X_a}\partial{X_a}e^{ipX}(z)}}
and
\eqn\grav{\eqalign{W(p)=\Omega^{ab}_m\lbrace
K\circ(\int{dz}{e^\phi}\lambda\partial{X_a}\partial{X_b}e^{ipX}(z)){\bar{L}}^m
({\bar{z}})
\cr
-{1\over2}
K\circ(\int{dz}{e^\phi}\lambda\partial{X^m}\partial{X_a}e^{ipX}(z)){\bar{L}}_b
+c.c.}}
with the $L$-operator given by (44).
As previously, BRST nontriviality conditions lead to
gauge transformations while  BRST-invariance constraints
on $U+W$ lead to linearized
equations of motion for the gauge fields.
Note that the frame field is an open string operator, similar to the
spin 3 vertex considered in our previous works ~{\spinself, \spinselff}. The
operator for $\Omega$ is, in turn, a closed string vertex operator
which structure is based on spin 3 operator combined with 
$L$-operator related to transvections in $AdS$. The leading order
contribution to beta-function of the sigma-model (67) with 
spin 3 operators (69), (70) stems from disc amplitudes 
and leads to the low-energy equations of motion:
\eqn\grav{\eqalign{
{1\over2}p_{\lbrack{m}}\Omega^{ab}_{n\rbrack}+\epsilon^{acd}
(e_{c{\lbrack}m}\wedge\Omega_{n{\rbrack}d}^b+\omega_{c{\lbrack}m}{E_{n{\rbrack}d}^b})=0
\cr
{1\over2}p_{\lbrack{m}}E^{ab}_{n\rbrack}+
\epsilon^{acd}
(\omega_{c{\lbrack}m}\wedge\Omega_{n{\rbrack}d}^b-{1\over{\rho^2}}
e_{c{\lbrack}m}{E_{n{\rbrack}d}^b})=0}}
which are the equations of motion for the higher spin part of
the Chern-Simons type theory ~{\hsafour, \hsafive, \hsasix,
\hsaseven, \soojongf, \henneaux, \campo, \gaber}:
\eqn\grav{\eqalign{
S=S(\Gamma_{+})-S(\Gamma_{-})\cr
S(\Gamma)\sim\int_{M_3}Tr(\Gamma\wedge{d}\Gamma+{2\over3}\Gamma\wedge
\Gamma\wedge\Gamma)}}

with the gauge fields $\Gamma$  taking values in 
higher spin algebra $hs(1,1)$ truncated 
to $sl(3,R)$  with $sl(3,R)$ components given by
\eqn\grav{\eqalign{
A^{a}_\pm=\omega^a\pm{1\over{\rho}}e^a
\cr
A^{ab}_\pm=\Omega^{ab}\pm{1\over{\rho}}E^{ab}}}

(while the $sl(2,R)$ truncation gives the equations
(66) for the MMSW gravity on AdS derived earlier in this paper
from string theory).
Extending these results to include higher spin components
of $hs(1,1)$ is a challenging and important problem, which requires
better understanding of vertex operators of higher ghost cohomologies 
~{\selfc}.
In the string theory context, the higher spin algebra $hs(1,1)$
should be realized as an operator algebra of the vertex operators living in 
higher order ghost cohomologies. 
 It would be particularly interesting to relate the asymptotic
$W_\infty$ symmetry of the Chern-Simons theory based on $hs(1,1)$,
discovered in  remarkable paper by Henneaux and S.-J. Rey
~{\soojongf}, to
internal symmetries of  {\it string field theory}
based on the action (72) with the higher spin operators
realizing hs(1,1) being the components of the string field $\Gamma$.
We hope to be able to elaborate on these ideas in future works.
To conclude, the sigma-model for connections and gauge fields
constructed in this work  provides a promising framework to
approach the unfolded dynamics of higher spin fields on $AdS$,
although understanding of vertex operator structure for
higher spin fields in frame-like description beyond $s=3$,
as well as of the underlying string field theory,
still needs to be developed.

\centerline{\bf Acknowledgements}

I would like to thank Robert De Mello Koch,
Soo-Jong Rey, Eugene Skvortsov and Misha Vasiliev
for interesting and stimulating discussions,
as well as for useful remarks and references.

\listrefs

\end